# Modeling Inner Boundary Values at 18 Solar Radii During Solar Quiet time for Global Three-dimensional Time-Dependent Magnetohydrodynamic Numerical Simulation

Chin-Chun Wu[1], Kan Liou[2], Simon Plunkett[1], Dennis Socker[1], Y.M. Wang[1], Brian Wood[1], S. T. Wu[3], Murray Dryer[4], and Christopher Kung[5]

[1] Naval Research Laboratory, Washington, DC 20375, USA
[2] Applied Physics Laboratory, Johns Hopkins University, Laurel, Maryland, USA
[3] CSPAR, University of Alabama, Huntsville, Alabama, USA (Deceased)
[4] Emeritus, NOAA, Boulder, Colorado, USA
[5] Engility Corporation, HPCMP PETTT, NRL, Code 5590, 4555 Overlook Ave, SW, Washington, DC 20375, USA

**27-September-2018**

**Abstract**

The solar wind speed plays a key role in the transport of coronal mass ejections (CME) out of the Sun and ultimately determines the arrival time of CME-driven shocks in the heliosphere. Here, we develop an empirical model of the solar wind parameters at the inner boundary (18 solar radii, *Rs*) used in our global, three-dimensional (3D) magnetohydrodynamic (MHD) model (G3DMHD) or other equivalent ones. The model takes solar magnetic field maps at 2.5 *Rs* (which is based on the Potential Field Source Surface, PFSS model) and interpolates the solar wind plasma and field out to 18 Rs using the algorithm of Wang and Sheeley [1990a]. A formula $V_{18Rs} = V_1 + V_2 f_s^\alpha$ is used to calculate the solar wind speed at 18 *Rs*, where $V_1$ is in a range of 150-350 km/s, $V_2$ is in the range of 250-500 km/s, and "$f_s$" is an expansion factor, which was derived from the Wang and Sheeley (WS) algorithm at 2.5 *Rs*. To estimate the solar wind density and temperature at 18 *Rs*, we assume an incompressible solar wind and a constant total pressure. The three free parameters are obtained by adjusting simulation results to match in-situ observations (*Wind*) for more than 54 combination of $V_1$, $V_2$ and α during a quiet solar wind interval, Carrington Rotation (CR) 2082. We found $V_{18Rs} = (150\pm50) + (500\pm100) f_s^{-0.4}$ km/s performs reasonably well in predicting solar wind parameters at 1 AU not just for CR 2082 but other quiet solar period. Comparing results from the present study with those from WSA [Arge *et al.* 2000; 2004] we conclude that i) Results of using $V_{18Rs}$ with the full rotation data (FR) as input to drive 3DMHD model is better than the results of WSA using FR, or daily updated. ii) When using a modified daily updated 4-day-advanced solar wind speed predictions WSA performs slightly better than our WSW-3DMHD. iii) When using $V_{18Rs}$ as input, 3DMHD model performs much better than the WSA formula. We argue the necessity of the extra angular width ($\theta_b$) parameter used in WSA.

# 1. Introduction

Coronal mass ejections (CMEs) are sudden eruptions of huge bubbles of coronal material into the interplanetary medium. When the structure moves into the solar wind, it is known as an interplanetary coronal mass ejection (ICME) [Dryer *et al.* 1994]. A fast-mode shock may result at the leading edge of the CME front [*e.g.*, Gosling *et al.,* 1975; Sheeley *et al*., 1982]. If any part of the shock arrives at Earth this knowledge can be used as a harbinger of geomagnetic activity. Thus the times of arrival of the shock and its following CME are important operational parameters. ICMEs are found to be associated with low-density magnetic clouds (hereafter MCs; Burlaga *et al.* 1981, 1982; Klein and Burlaga, 1982). About 30% of ICMEs are MCs [*e.g.* Gosling et al. 1990; Wu and Lepping 2007]. Observations have shown that a large percentage of MCs/ICMEs lead to magnetic storms [*e.g*., Wu and Lepping, 2002; Huttunen *et al*. 2005; Zhang *et al*. 2007]. This is probably due to the large fluctuating magnetic field behind the shock and the large, smooth, and long-duration cloud field inside the cloud that favor magnetic merging. Indeed, many extremely large geomagnetic storms are associated with CME events [*e.g*., Zhang *et al.* 2007; Wu *et al.* 2013]. Therefore, accurate and timely forecasting the arrival of these events becomes an important imperative in order to protect expensive space assets and astronauts, and to minimize communications interruptions.

Numerical time-dependent, three-dimensional (3D), magnetohydrodynamic (MHD) models are theoretically capable of predicting solar wind parameters from the Sun to the Earth. Han *et al.* [1988] developed the first time-dependent, 3-D, MHD simulation model. The model has been used to study interplanetary (IP) shock evolution from 18 solar radii (Rs) or 0.1 AU to the Earth [*e.g.,* Han *et al.* 1988; Detman *et al.* 1991; Dryer *et al.* 1997; Wu and Dryer, 1997; Wu *et al.* 1996; 2005]. We will refer to this model as Han's code hereafter. Han's code has also been used previously to study (i) interplanetary magnetic field (IMF) draping around plasmoids in the solar wind [Detman *et al.,* 1991]; (ii) IMF changes at 1 AU as a consequence of an interaction with a heliospheric current/plasma sheet (HCS/HPS) [Wu *et al.* 1996; Wu and Dryer, 1997]; and (iii) shock arrival time at the Earth [Wu *et al.* 2005]. Several early examples include evolution of a shock which was driven by a CME that occurred on 14 April 1994 and its propagation to the Earth and at ~4 AU [Dryer *et al.* 1997]. Pressure pulses have also been utilized at lower boundaries to mimic solar events to study the evolution of solar transient disturbances (*e.g.*, shocks, plasma clouds, and magnetic flux ropes) by other groups [*e.g.*, Odstrcil and Pizzo, 1999a,b; Groth *et al.* 2000; Hayashi *et al.* 2011; Manchester *et al.* 2004; Vandas *et al.* 2002; Luguz *et al.* 2011; Shen *et al.* 2011].

Potential field source-surface models are often used to derive ambient solar wind parameters at the inner boundary of heliospheric MHD models [*e.g.* Usmanov 1993; Manchester *et al.* 2004; Odstrcil *et al.* 2005; Detman *et al.* 2006; Luguz *et al.* 2011; Shen *et al.* 2011; Wu *et al.* 2007a,b]. Han's code and the Hakamada, Akasofu and Fry (HAF) code [Fry *et al.* 2001] were merged as a hybrid model (HAF+3DMHD) to simulate realistic solar wind structures from 2.5 Rs to the Earth environment and beyond [Liou *et al.* 2014; Wood *et al.* 2011, 2012; Wu *et al.* 2007a,b, 2011, 2012, 2016a,b]. The combined HAF+3DMHD model is capable of simulating extremely fast CME events, for example, the fastest recorded CME that erupted on 23 July 2012 with a shock speed ($V_S$) faster than 3000 km/s [Liou *et al.,* 2014]. It is also capable of modeling the evolution and interaction of multiple CMEs [*e.g.*, Wu *et al.* 2012; Wu *et al.* 2016b; S.T.Wu *et al.* 2014].

Using 22 years of flux-tube expansion factor ($f_s$, which was derived near the Sun), Wang and Sheeley [1990a,b] constructed an empirical model that is capable of estimating daily characteristic solar wind speed at the Earth (WS model). According to the values of $f_s$, they cataloged solar wind speed into six values: 700 km/s ($f_s < 3.5$), 600 km/s ($3.5 < f_s < 9$), 500 km/s ($9 < f_s < 18$), 400 km/s ($18 < f_s < 54$), and 330 km/s ($54 < f_s$), respectively. The WS $v$-$f_s$ relationship is based upon the empirical correlation found between the solar wind velocities observed near the Earth with the corresponding $f_s$ values determined at the source surface. These two quantities are linked by the time required for the radially propagating solar wind (assumed to be flowing at constant velocity) to traverse from Sun to Earth.

The velocity profile produced by the WS velocity scheme is discrete. Therefore, the WS velocity relationship cannot be used as input for the global MHD simulation. Arge and Pizzo [2000] (AP) made a number of modifications to the basic technique of the WS model. The AP $v$-$f_s$ relationship is a continuous empirical function that related magnetic expansion factor to solar wind velocity at the source surface. The AP $v$-$f_s$ relationship used daily updated synoptic maps instead of full-rotation maps. Both WS and AP $v$-$f_s$ relationship use solar wind speed at the Lagrangian point 1 (L1) to trace back to the solar source surface. Solar wind speed is highly non-uniform near the Sun.

The ambient (pre-existing background) solar wind speed is known to affect the acceleration and deceleration of CMEs [*e.g.*, Gopalswamy *et al.* 2009; Wu *et al.* 2006]. Time-dependent, 3D MHD simulations also show that the background solar wind can affect the arrival time of shock events with slow propagation speed ($V_{Shock} < 100$ km/s) but not the shock events with fast propagation speed [*e.g.,* Wu *et al.* 2005]. Current 3D global MHD models often overestimate the background solar wind speed at the inner boundaries, *e.g.*, works performed by Wu *et al.* [2016a,b] with the HAF+3DMHD model and by Yu *et al.* [2015] with the ENLIL model using solar wind velocity deriving from the interplanetary scintillation (IPS) remote-sensing method. In their simulation using the ENLIL model, Yu *et al.* had to reduce the solar wind speed input at 0.1

AU by ~20% to get the right IP shock arrival time at the Earth. For space weather forecasting purposes, it is important to be able to obtain the correct initial solar wind speed as a simulation input. Therefore we are motivated to develop a scheme of providing solar wind velocity at the inner boundary (18Rs) for three-dimensional, time-dependent MHD simulation model that could produce realistic background solar wind condition at the Earth. The remaining sections of the paper are organized as follows. We will describe the numerical simulation in Section 2. In Section 3, we demonstrate the methodology. Tuning, including validation and discussion of simulation results (*i.e*., parameter tuning for 1 AU solar wind speed), is described in Section 3. Conclusions and Remarks are given in Section 4.

**2. Global Three-Dimensional MHD Simulation Model (G3DMHD)**

2.1 **3-D MHD simulation model**

The fully 3-D, time-dependent MHD simulation code [Han *et al*., 1977, 1988] was used to propagate solar wind parameters at the inner boundary to 1 AU to compare with *in situ* measurements. The MHD model solves a set of ideal-MHD equations using an extension scheme of the two-step Lax-Wendroff finite difference methods [Lax and Wendroff, 1960]. An ideal MHD fluid is assumed in the Han model, which solves the basic conservation laws (mass, momentum, and energy) as shown in Equations (1) - (3) with the induction equation (Equation 4) to take into account the nonlinear interaction between plasma flow and magnetic field.

$$\frac{D\rho}{Dt} + \rho \nabla \bullet \mathbf{V} = 0 \qquad (1)$$

$$\rho \frac{D\mathbf{V}}{Dt} = -\nabla p + \frac{1}{\mu_o}(\nabla \times \mathbf{B}) \times \mathbf{B} - \rho \frac{GM(r)}{r^2}\hat{\mathbf{r}} \qquad (2)$$

$$\frac{\partial}{\partial t}[\rho e + \frac{1}{2}\rho |V|^2 + \frac{|B|^2}{2\mu_o}] + \nabla \bullet [V\{\rho e + \frac{1}{2}\rho |V|^2 + p\} + \frac{B \times (V \times B)}{\mu_o}] = -v \bullet \rho \frac{GM(r)}{r^2}\hat{r}$$
(3)

$$\frac{\partial B}{\partial t} = \nabla \times (V \times B)$$
(4)

where *t, r, ρ, V, B, p, e* are time, radius, density, velocity, magnetic field, thermal pressure, and internal energy. The internal energy, $e \equiv p/[(\gamma-1)\rho]$. Additional symbols γ, $M_s$, *G*, $\mu_o$ are the polytropic index, the solar mass, the gravitational constant, and the magnetic permeability in vacuum. γ = 5/3 is used for this study since it has been shown to be a good value to use for in-situ solar wind data at 1 AU [*e.g.* Wu *et al.*, 2011; Liou *et al.*, 2014]. The MHD governing equations are cast in uniform, spherical grids. The computational domain for the 3-D MHD simulation is a sun-centered spherical coordinate system (r, θ, φ) oriented on the ecliptic plane. Earth is located at *r = 215 $R_s$, θ = 0°*, and *φ = 180°*. The domain covers *-87.5° ≤ θ ≤ 87.5°; 0° ≤ φ ≤ 360°; 18 $R_s$ ≤ r ≤ 345 $R_s$*. An open boundary condition at both θ = 87.5°, θ = -87.5°, and r = 245 $R_s$ are used so there are no reflective disturbances. A constant grid size of *Δr = 3 $R_s$, Δθ=5°*, and *Δφ=5°* is used which results in 110×36×72 grid sets.

## 2.2 Inner Boundary Data Set Up

The system is driven by a time series of photospheric magnetic maps composed from daily solar photospheric magnetograms (http://wso.stanford.edu). The WS model uses the observed line-of-sight magnetic field at the photosphere extrapolated to 2.5 $R_s$ [*e.g.*, Wang and Sheeley, 1992]. The inner boundary of the 3-D MHD model is at an adjustable location, typically beyond the critical points at 18 solar radii ($R_s$). The conservation of magnetic flux (*r $B_r^2$* = constant) is used to derive magnetic field at 18 $R_s$. Conservation of the flux tube *r $B_r^2$* = constant is assumed to set up spacing variation (*i.e.* grid size) in both $\theta$- and $\phi$-direction. A formula $V_r = V_1 + V_2 f_s^{\alpha}$ (units in km/s) is used to compute $V_r$ at 18 $R_s$, where $V_1$ is a constant ranging from 150 to 350, $V_2$ is also a constant ranging from 250 to 500, $f_s$ is the expansion factor [Wang and Sheeley, 1990a, b, 1992], and $\alpha$ is the exponent of the expansion factor. This is similar to the work done by Arge [2004]. Conservation of mass, $\rho V = \rho_o V_o$ = constant, is used to compute the solar wind density at 18 $R_s$, where $\rho_o$ is $2.35 \times 10^{-9}$ kg/km$^3$ and $V_o$ is the average of $V_r$ at 18 $R_s$. We further assume that the total pressure is constant along the stream line (Bernoulli's principle). The equation $\rho (RT + v^2/2) = \rho_o (RT_o + v_o^2/2)$ = constant is used to compute temperature at 18 $R_s$, where $T_o = 1.5 \times 10^6$ °K is used at 18 $R_s$.

**2.3 Selection of Study period**

The occurrence frequency of CMEs ranges from ~0.6/day to ~4/day [*e.g.,* Wu, Lepping, and Gopalswamy, 2006] or to ~6/day [Wang and Colaninno, 2014; Hess and Colaninno, 2017; Vourlidas *et al.* 2017], depending on the phase of the solar cycle. When a CME/ICME/Shock propagates from the Sun to the Earth, the solar wind can vary a lot, depending on the size/speed of the CME. For constructing a global MHD simulation model, a quiet solar wind period is a better choice to test the model. Therefore, we picked a quiet period (*i.e.* sunspot number, SSN is small) during which the occurrence frequency of CMEs is also low. The value of the 13-month smoothed monthly total SSN is 3.4 in April-May 2009 (http://www.sidc.be/silso/datafiles). The average of the monthly CME occurrence was ~108.7 during 1996-2015, and about 60 and 47 CME were observed in April and May 2009, respectively (CME data was obtained from website https://cdaw.gsfc.nasa.gov/CME_list/). No MCs were observed during April - May 2009 [Lepping *et al.* 2015]. In addition, no magnetic cloud-like structure was found in 2009 [Wu and Lepping, 2015]. Therefore, Carrington Rotation (CR) 2082 (April 5 to May 3, 2009) was chosen to test our new solar wind speed scheme/model under quiet conditions.

Figure 1 shows the background (co-rotating "steady state") solar wind radial speed ($V_r$) on the surface plane at 18 and 216 $R_s$ at 02:00UT on 3 April 2009. These values are calculated using $Vr = 150 + 250 f_s^{-0.4}$ (Fig. 1a-b) and $Vr = 150 + 500 f_s^{-0.4}$ (Fig. 1c-d). The solar wind speed is faster at 216$Rs$ (see Fig. 1b and 1d) than that at 18$R_s$ (see Fig. 1a and 1c). Overall, Figure 1 clearly shows that solar wind speed using the formula $Vr = 150 + 500 f_s^{-0.4}$ is faster than that obtained by using the formula $Vr = 150 + 250 f_s^{-0.4}$.

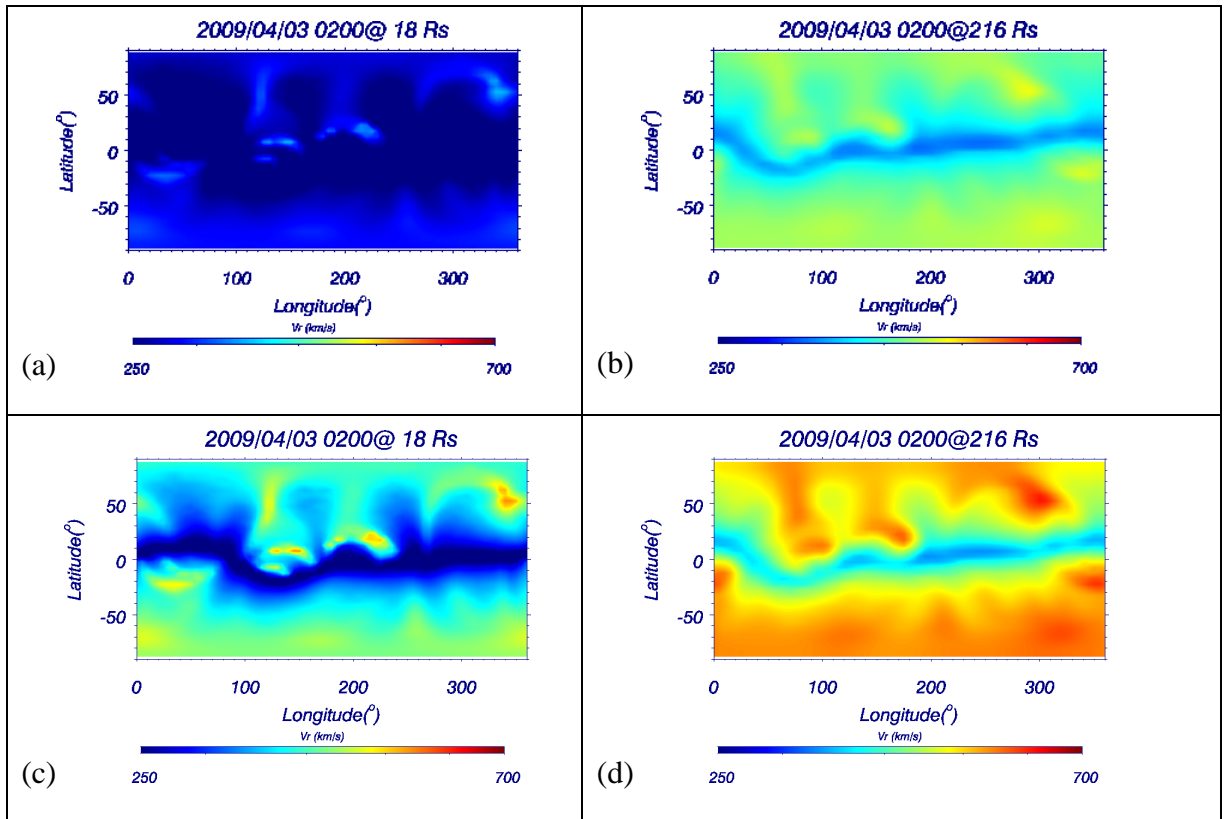

Figure 1. Background (corotating "steady state") solar wind condition in the plane at (a,c) 18 and (b,d) 216 Rs on 4 April 2009, 15:00UT by using velocity formula, $Vr = 150 + 250 f_s^{-0.4}$ (Fig. 1a-b), and $Vr = 150 + 500 f_s^{-0.4}$ (Fig. 1c-d).

### 2.4 Setting up co-rotating steady state solar wind

The governing MHD equations are described in the inertial frame, thus, the solar sidereal rotation vector, ~Ω, does not appear in the governing equations. Instead of using the rotating frame as the reference coordinate system, we assume that the distribution map of the inner boundary values at 18 $Rs$ moves longitudinally at the solar sidereal rotation rate in the inertial system. We set the solar rotation rate $|\Omega|$ to be 360 degrees per 27.27 days. On 2 April 2009, the Earth was located at a latitude of south 6.6° (S6.6°) with respect to the solar equator. Figure 2 shows the velocity profile at 2.5 south (S2.5°) of the solar-equatorial plane using the formula, $Vr = 150 + 300 f_s^{-0.4}$ for velocity profile (distribution) at 18 Rs. Figure 2a shows that the solar wind has no spiral feature initially. Everything goes out radially. Figure 2 shows that the solar wind takes about 4 days to reach a spiral configuration at 1 AU (Figure 2d), and about 6 days to reach a spiral configuration for the entire simulation domain (Figure 2e).

Figure 1 shows clearly that the velocity profile in both $\theta$-, and $\phi$- directions are non-uniform. The flow speed values are larger in the high-latitude regions than those in the low-latitude regions. Figures 3a-3d and 3e-3h show the solar wind speed and density on surfaces of different angular cones that are centered at the Sun's center. These conical angles are at 22.5°N (north, representative of a response in the northern heliosphere), 7.5°N, 7.5°S (close to Earth's latitude in the solar equatorial coordinate system), and 22.5°S (south, representative of a response in the southern heliosphere). Figures 3i-3m show the solar wind speed at different longitudinal meridian planes: 90°E (East, Fig.3i), 45°E (Fig.3j), 0°W (west, Fig.3k, Sun-Earth-line direction), 45°W (Fig.3l), and 90°W (Fig.3m).

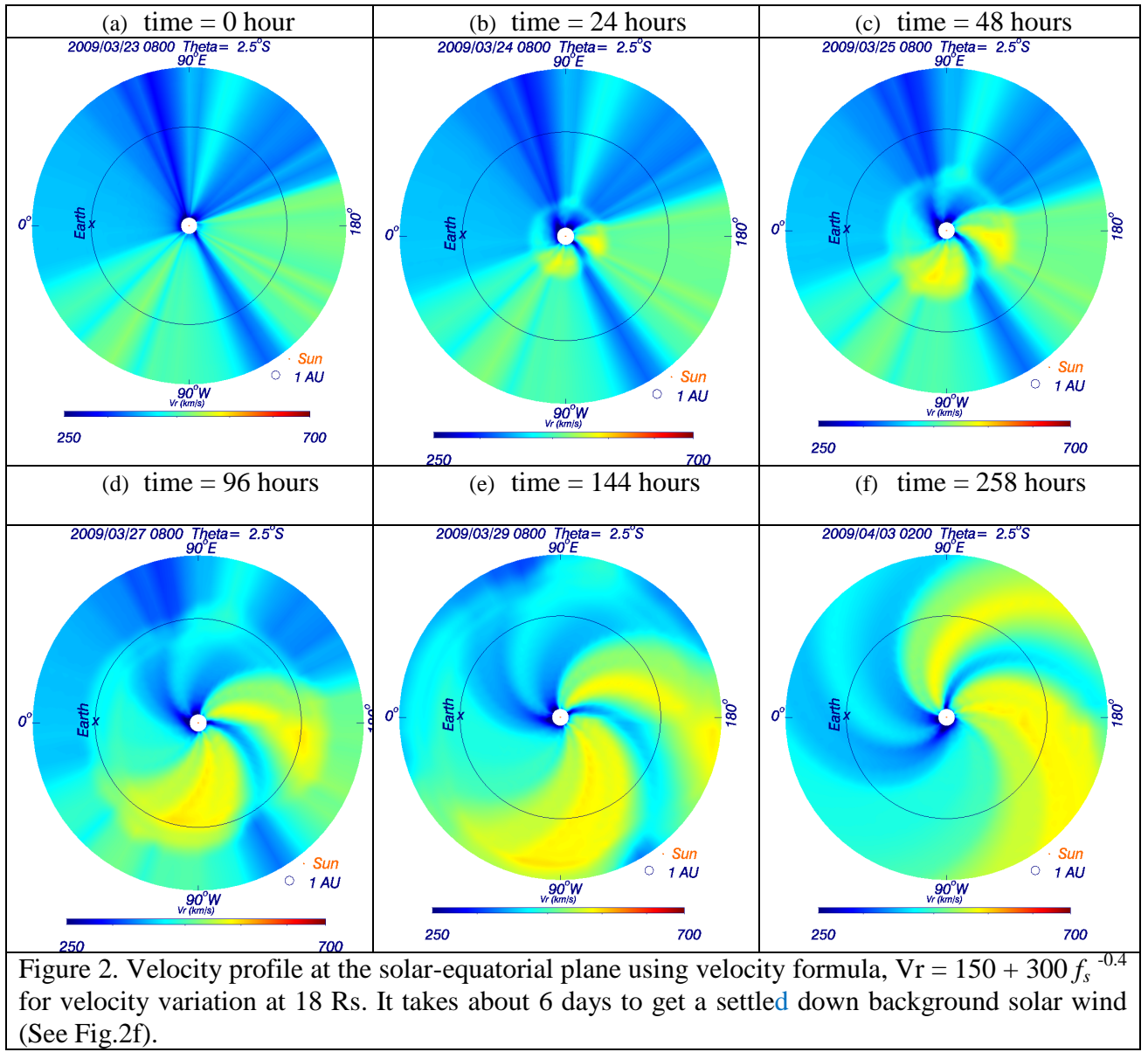

Figure 2. Velocity profile at the solar-equatorial plane using velocity formula, $V_r = 150 + 300\, f_s^{-0.4}$ for velocity variation at 18 Rs. It takes about 6 days to get a settled down background solar wind (See Fig.2f).

The solar wind speed profiles are highly non-uniform. For example, i) solar wind speed was lower at the inner boundary (*i.e.* 18 $R_s$) than it was at 1 AU (*i.e.* 215 $R_s$; ii) solar wind speed was higher in the southern hemisphere than that in the northern hemisphere; iii) the highest speed stream was located near 180ºW in the southern hemisphere, but near the 5ºW in the northern hemisphere; iv) solar wind speed was slower near the equator than in the higher latitude regions (See Fig. 3a-3d, and 3i-3m).

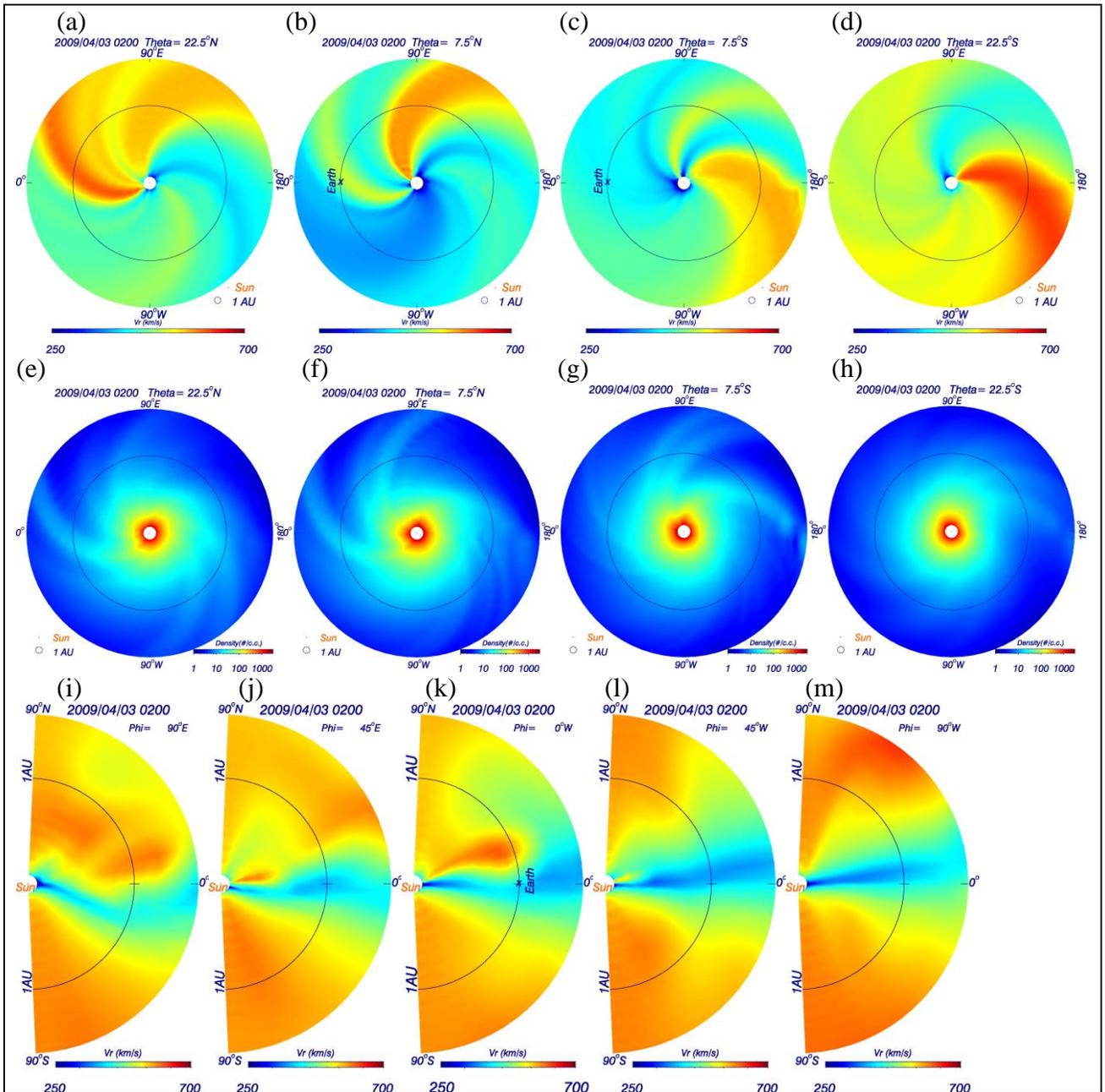

**Figure 3.** Solar wind speed (a-d) and density (e-h) on surfaces of different angular cones that are centered at the Sun's center. These conical angles are at 22.5ºN (north, representative of a response in the northern heliosphere), 7.5ºN, 7.5ºS (close to Earth's latitude in the solar equatorial coordinate system), and 22.5ºS (south, representative of a response in the southern heliosphere). Figures 3i-3m show the solar wind speed at different longitudinal meridian plane: 90ºE (East, Fig.3i), 45ºE (Fig.3j), 0ºW (west, 3k), 45ºW (Fig.3l), and 90ºW (Fig.3m).

## 3. Validation of Simulation Results and Discussion

### 3.1 Effect of $V_1$ and $V_2$ on the solar wind profile

The initial background solar wind condition is initialized by an input solar wind velocity profile of $V_1+V_2 f_s^{-0.4}$ km/s that is used to specify the velocity distribution at the inner boundary sphere at 18 $R_s$ for this study. In which, $V_1$ is the baseline solar wind speed, and $V_2$ is the amplitude of velocity above the baseline. In this study, we mainly concentrate on the testing of solar wind speed by using the WS expansion factor based on the solar magnetogram measurements from the Wilcox Solar Observatory (WSO) together with the potential field model to construct the initial solar wind speed profile at 18 $R_s$, in order to find the best parameters for constructing the background solar wind speed. Fifty-four cases (combinations of $V_1$ and $V_2$) were simulated. $V_1$ was in a range of 150-350 km/s, and $\Delta V_1 = 25$ km/s was used. $V_2$ was in a range of 250-600 km/s, and $\Delta V_2 = 50$ km/s was used. Fifty-four cases with different combinations of nine different values of $V_1$ (150, 175, 200, 225, 250, 275, 300, 325, 350 km/s) and six values of $V_2$ (250, 300, 350, 400, 450, 500, 550, 600 km/s) were simulated. Note that the exponent α held constant at a value of -0.4 for each of these cases.

Time profiles of the solar wind speed at the Earth for the period between March 30 and April 27, 2009 for 72 cases are presented in Figure 4. Red-dotted lines represent *in-situ* solar wind speed (from the OMNI database, https://omniweb.gsfc.nasa.gov), and black-solid lines represent simulated solar wind speed using the G3DMHD (G3DMHD: WSW+3DMHD) model. From top to bottom (Panels 1 to 9): $V_1$ was 150, 175, 200, 225, 250, 275, 300, 325, and 350 km/s, respectively. From left to right: $V_2$ was 250, 300, 350, 400, 450, 500, 550 and 600, respectively. The correlation coefficient (cc), the mean absolute percentage error (MAPE) [≡ (100/N x ∑ |$V_{Wind}$ − $V_{G3DMHD}$|/$V_{Wind}$|], and standard deviation (σ) for hourly averaged observation vs.

simulation (G3DMHD) are marked on the top of each panel (from left to right of each panel). The values of the cc, MAPE, and σ were in a range of 0.56-0.80, 7% to 49%, and 29-186 km/s, respectively

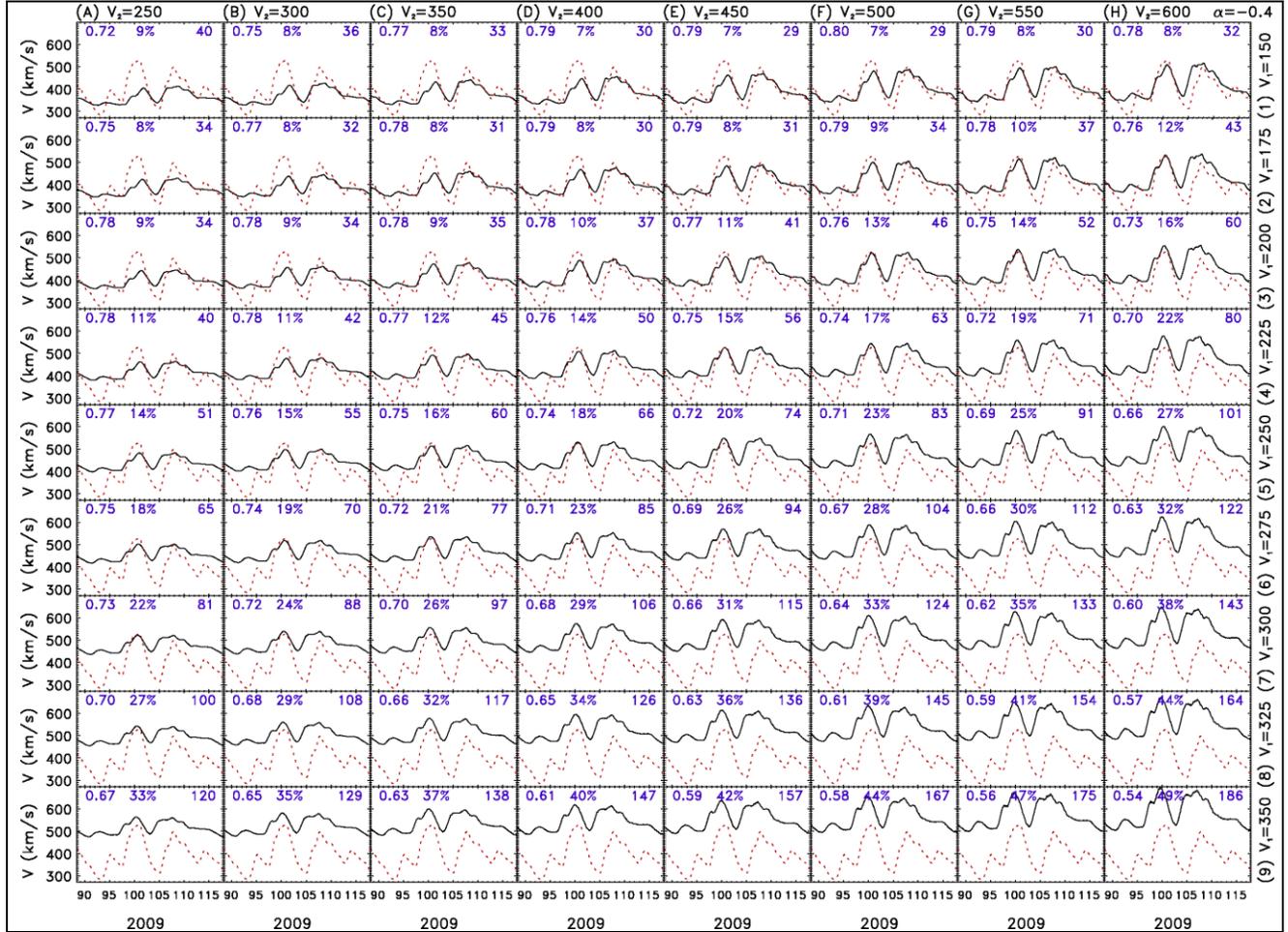

Figure 4. Variation of solar wind speed at L1 during March-April 2009. Red-dotted and Black-solid lines represent observation (OMNI) and H3dMHD simulation results. Solar wind speed was constructed by using speed formula, $V_{18Rs} = V_1 + V_2 f_s^{-0.4}$ (km/s). $V_1$ ranges between 150 and 350 (top to bottom panels 1-9: $V_1$ was 150, 175, 200, 225, 250, 275, 300, 325, and 350, respectively). $V_2$ ranges between 250 to 600 km/s (left to right panels A-H: $V_2$ was 250, 350, 400, 450, 500, 550, and 600 km/s, respectively). $f_s$ is the expansion factor which was derived by using Wang and Sheeley model [1990].

Overall, during March 20-April 27, 2009, Case 1F ($Vr = 150 + 500f_s^{-0.4}$) has the best correlation coefficient (=0.80) and also has a very low value of MAPE (= 7%). Case 1E ($Vr = 150 + 450f_s^{-0.4}$) also has a good fit, except the cc is 0.01 less than Case 1F. Other cases also have a high value of cc and a low MAPE, but the trend is not as good as Case 1F, *i.e.* matched velocity profile for both velocity in the minimum ($V_{min}$, minimum velocity) and maximum ($V_{max}$, maximum velocity). For example, $V_{max}$ is far off the observation in either Case 3A ($Vr = 200 + 250f_s^{-0.4}$, cc =0.78, σ=34), or Case 2C ($Vr = 200 + 350 f_s^{-0.4}$, cc =0.78, MAPE =8%).

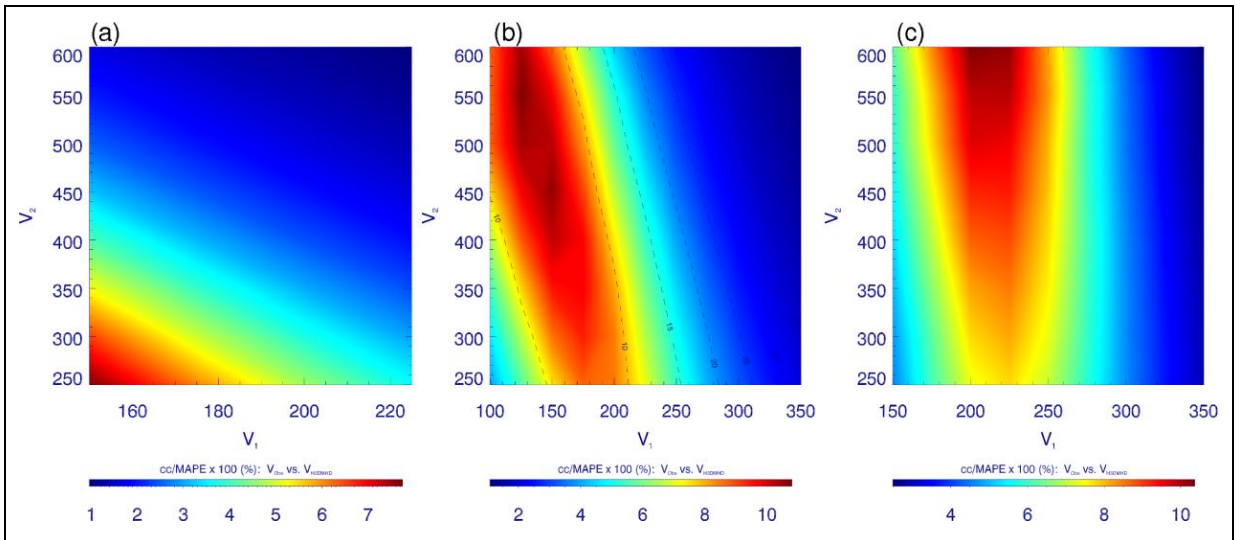

Figure 5. Ratio of Correlation coefficient (cc) over MAPE for different $V_1$s' (ranges between 150 and 350) and $V_2$s' (ranges between 250 and 600) for CR2082. Colors represent the ratio of "cc divided by MAPE" x 100 (%). Blue-dashed-contours are MAPE between observations and simulation results.

Using the velocity formula $V_{18Rs} = V_1 + V_2 f_s^{-0.4}$ to construct solar wind speed (see Figure 4) at the inner boundary, two major trends of solar wind speed near the Earth are identified: (i) the baseline solar wind speed was low if a low value of $V_1$ is used. (ii) The peak solar wind speed ($V_{peak}$) is high if a large value of $V_2$ is used. The trend of the speed variation is similar between the observations and the simulations for cases with $V_1$ less than 225 km/s (Panels 1-4). For cases with a high value of $V_1$ (*i.e.*, $V_1 > 250$ km/s), the simulated speed baselines were much higher than observed (See panels 6-9 of Fig.4). Overall the equation $V_{18Rs} = (125\pm25) + (500\pm100) f_s^{-0.4}$ is a good fit to background solar wind at 1 AU.

To explore wider parameter regimes, we have performed simulations for α = 0.2 and 0.6. Since the best correlation coefficient does not always coincide with the smallest MAPE, we define a new metric, cc/MAPE, which is defined as the ratio of cc to MAPE. Based on the new metric, a better cc and smaller MAPE lead to a larger cc/MAPE. Figures 5(a) - 5(c) show the contours of cc/MAPE for α = 0.2, 0.4, and 0.6, respectively. For α = 0.2, the best $V_2$ is smaller than 250 km/s and the best $V_1$ is smaller than 150 km/s, all outside the $V_1$ and $V_2$ parameter regions considered here. On the other hand, for α = 0.6, while the best $V_1$ is 210 km/s and within its parameter regime, the best $V_2$ is greater than 600 km/s, outside the $V_2$ parameter regimes. Let's consider the WS formula, $V = V_1 + V_2 \cdot f_s^{-\alpha}$, it implies that $V_1$ is the minimum wind speed at 18 Rs (when $f_s \gg 1$, $V \approx V_1$) and $V_2$ is the difference between the fastest and slowest wind speed at 18 Rs (when $f_s \to 1$ and $V \to V_1 + V_2$). If we impose the minimum and maximum solar wind speed as 200 and 800 km/s at 18 Rs, the present study suggests that α must be greater than 0.2 and less than 0.6 because both limits introduce values of $V_1$ and $V_2$ outside this speed range. When α = 0.4, the preference values for $V_1$ and $V_2$ fall within their respected range. Of course, α can be 0.3 or 0.5 or any value between 0.2 and 0.6. However, we do not believe these values will introduce significant differences in terms of their prediction performance. There is a wide acceptable region for the $V_1$ and $V_2$ parameters in Figure 5(b). We will demonstrate next that a middle value of the $V_1$ and $V_2$, *e.g*, 150 and 500 km/s, can yield in general better results when other solar wind parameters are considered.

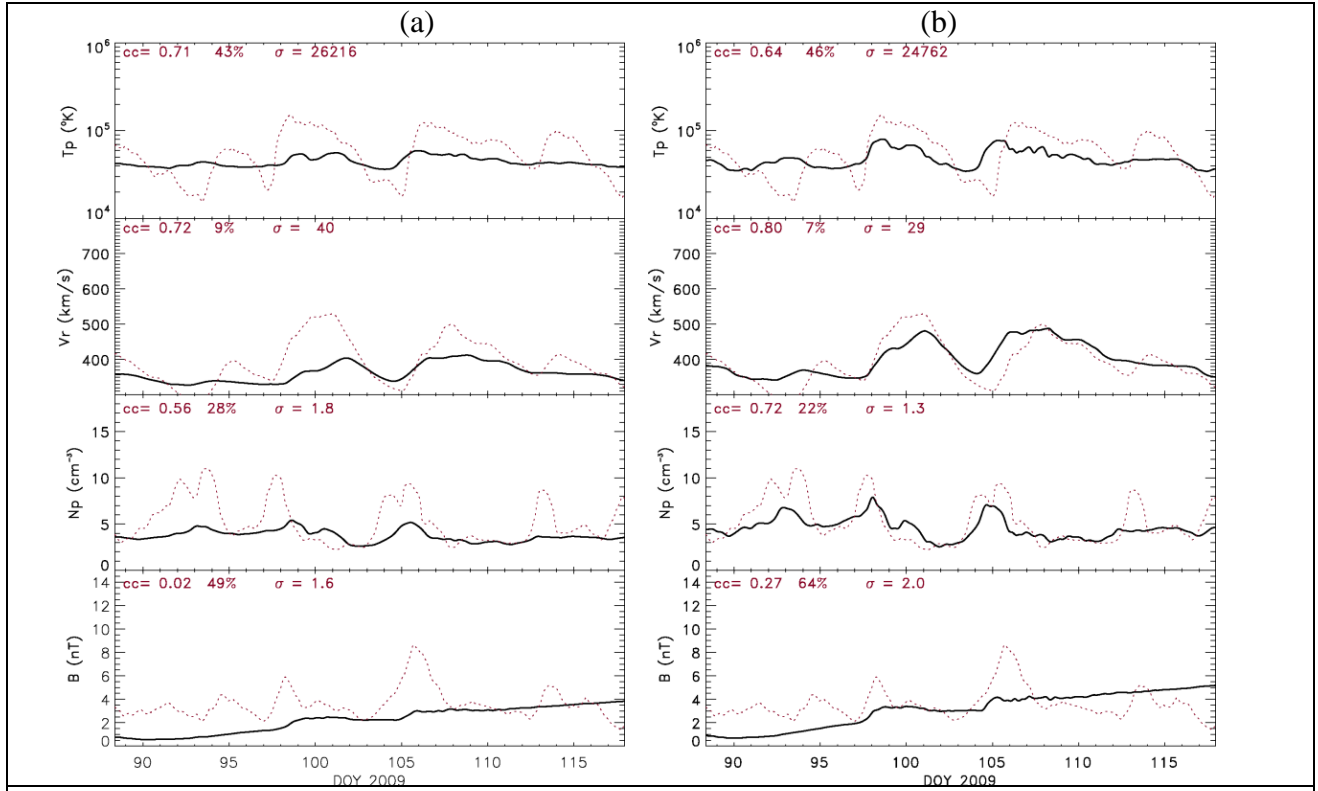

Figure 6. Comparison of the simulated background solar wind for H3DMHD (black-solid-lines, at S2.5°) vs. observation (red-dotted-lines). (a) $Vr=150+250 f_s^{-0.4}$ was used to construct solar wind speed at 18 Rs. (b) $Vr=150+500 f_s^{-0.4}$ was used to construct solar wind speed at 18 Rs.

We first draw attention to the comparison of the simulation results with the *in-situ* observations at Earth in Figure 6. $Vr=150+250 f_s^{-0.4}$ and $Vr=150+500 f_s^{-0.4}$ are used to produce the background solar wind in Figures 6a and 6b, respectively. Apparently these two parameter pairs fall within the large acceptable parameter range as shown in Figure 5(b). The time resolution of the observations is ≈1.5 minutes. The time resolution of simulated solar wind is in a range of 1-15 minutes, which depends on the simulated solar wind condition. Both data sets were interpolated into hourly resolution. Validation of our simulation results was done by comparing solar wind plasma and field parameters with *in situ* measurements at 1 AU (*e.g,* made by *Wind* or *ACE* spacecraft, or *OMNI* data set). Figure 6 shows a comparison of the solar wind parameters from G3DMHD simulations and observations during March 30 - April 27, 2009 for Cases 1A and 1F. Panels from top to bottom show the time profile of solar wind temperature (*Tp*, units in

°K), velocity in r-direction (Vr, units in km/s), density (Np, units in cm$^{-3}$), and magnitude of interplanetary magnetic field (B, units in nT). Earth was orbiting between 6.7° and 5.0° below solar equatorial plane (or S6.7° and S5.0°).

For the Case 1A, the averages of ambient solar wind parameters <Tp>, <Vr>, and <Np> were under-estimated by ~28%, 7%, and 28%, respectively (see Fig.6a); but the average of total magnetic field, <B> was over-estimated by 15%. The cc's for simulation vs. observation were 0.71, 0.72, 0.56, 0.02 for Tp, Vr, Np, and B, respectively. For the Case 1F, ambient solar wind <Tp> and <Np> were under-estimated by 22% (-22%) and 12% (-12%), respectively (See Fig. 6b); but <Vr> and <B> were over-estimated by 1% and 37%. The cc's for simulation vs. observation are 0.63, 0.79, 0.73, and 0.28 for Tp, Vr, Np, and B, respectively. Overall, the results for Case 1F are better than that for Case 1A.

## 3.2 Validation of the good fit formula, $V_{GF} = 150 + 500 f_s^{-0.4}$

Based on the results of Figures 4 and 5, $V = 150 + 500 f_s^{-0.4}$ provided one of the good fit results for CR2082. $V = 150 + 500 f_s^{-0.4}$ will be referred as $V_{GF}$ hereafter. The $V_{GF}$ empirical formula used in the study is similar (nearly identical) to the Equation (4), $v(f_s) = 267.5 + [410/f_s^{0.4}]$ (referred to $V_{AP}$ hereafter) that was used by Arge and Pizzo [2000]. The value of $V_1$ is lower for $V_{GF}$ than that for $V_{AP}$, but $V_2$ is higher for $V_{GF}$ than that for $V_{AP}$. Note that the $V_{AP}$ formula was used to estimate solar wind speed at 1 AU, but the $V_{GF}$ formula was used to estimate solar wind speed at 18 $R_s$ (~0.1 AU). This may cause the differences between $V_{AP}$ and $V_{GF}$.

Areg and Pizzo [2000] used $V_{AP}$ and three different source surface maps: (i) the full rotation (FR), (ii) daily updated (DU), and (iii) modified daily updated (MDU) 4-day-advanced solar wind speed predictions with 9-hour-averaged *Wind* velocity observation for CR1899. The correlation coefficients for observations vs. predictions are 0.678, 0.793, and 0.813 for using FR, DU, and MDU data sets, respectively (see Figure 4 in Arge and Pizzo [2000]).

In order to evaluate our $V_{GF}$ ($V = 150 + 500 f_s^{-0.4}$) formula, we used $V_{GF}$ to simulate the solar wind condition for CR1899 during 6 August – 3 September, 1995. A comparison of the full rotation G3DMHD/simulated solar wind *V* (top panel), *Np* (second panel from top), *Tp* (third panel from top), and *B* (bottom panel) with 9-hour-averaged *Wind* spacecraft solar wind observations (red dotted lines) are shown in Figure 7. Values of cc, MAPE, and σ are 0.803, 12.4%, and 49.4 km/s, respectively. Our result is better than AP's results using the full rotation (FR) data or the daily updated (DU) data. However, the cc is slightly less than AP's results of using modified daily updated (MUD) 4-day-advanced solar wind speed. Values of the cc are 0.48, 0.63, and -0.04 for *Np*, *Tp*, and B, respectively. Values of the MAPE are 0.01, 0.336, and 0.439 for *Np*, *Tp*, and *B*, respectively. We have to stress that a better linear correlation is not necessary a better fit. This is the reason that we use the value of cc/MAPE (See Figure 5) to evaluate the fit.

In this study we have carefully selected a period without any solar disturbance, and used about 72 different combinations of simple velocity empirical formula to find a best fit formula in the solar quiet time, *i.e.* in 2009 which is in the beginning of solar cycle 24. In the above paragraph, we demonstrated that the simple formula, $V_{GF}$ is also valid in 1995 which is in the end of solar cycle 22. One may argue about the capability of the $V_{GF}$ for other periods of time.

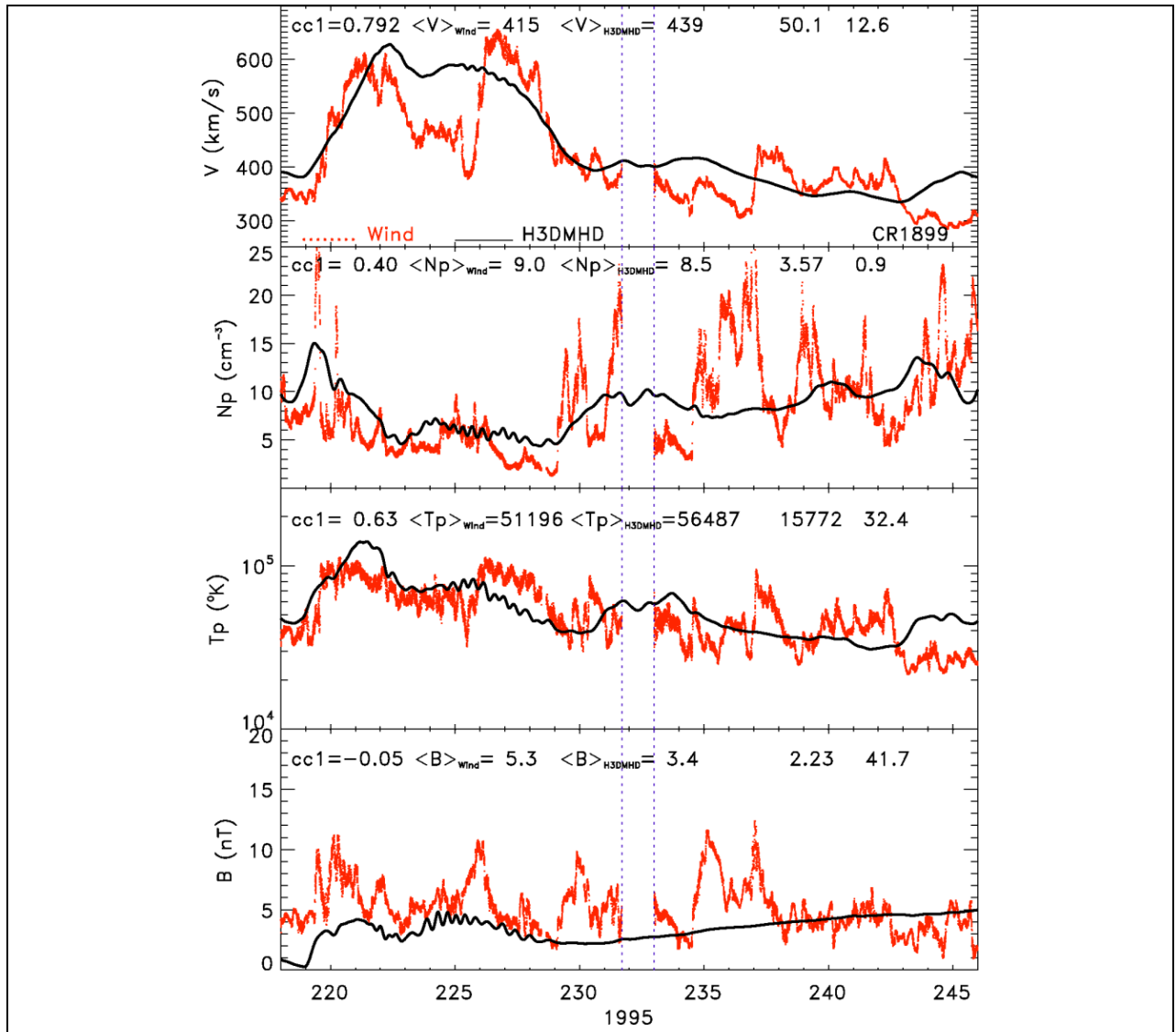

Figure 7. Comparison of solar wind speed, density, temperature, and temperature from the *WIND* spacecraft (red-dotted lines) with G3DMHD prediction (black solid lines) for CR1899 (during 6 August - 2 September 1995). A data gap of *Wind* was marked between two blue vertical dotted lines.

Riley *et al.* [2001] used a $\theta_b$ parameter, in addition to $f_s$, to empirically specify solar wind speed near the Sun for a number of years, where $\theta_b$ is the minimum angular separation (at the photosphere) between an open field foot point and its nearest coronal hole boundary introduced by Arge *et al.* [2003]. Their predicted velocity for CR 1921-1923 was shown in the Fig. 3 of Arge *et al.* [2004]. However, their prediction for CR1922 during the three-day period (May 8-11, 1997) was not correct. The WSA model predicted a fast stream during these three days. They claimed that using higher resolution maps may help to reduce some of these problems. In addition, WSA also made a false prediction of two high-speed streams during April 25-30, 1997. A high-speed stream observed by *Wind* during April 10-15 (in CR1921) was also missing from the WSA prediction. The stream during April 10-15 was caused by the crossing of an interplanetary coronal mass ejection (ICME), presumably associated with a CME that occurred on April 7 (Webb *et al.* 2000; Arge *et al.* 2004).

In order to explore the capability of $V_{GF}$ formula for predicting the background solar wind, we further consider the following three periods of solar rotation: CR1921, CR1922, and CR1923. The comparison of the G3DMHD simulated solar wind speed with the *Wind* in-situ solar wind speed is shown in Figure 8. The relationship between the observation and simulation is reasonably acceptable for the periods of CR1921 and CR1923, with a MAPE value of 14.9% and 17.1%, respectively. The performance is clearly much better for CR1922 (cc=0.80, MAPE = 11.6%). G3DMHD correctly predicted the two fast streams during April 30 – May 03, and May 15-18 (see middle panel of Figure 8). Furthermore, G3DMHD did not make the false prediction for the period of May 8-11, 1997 as made by Arge *et al.* [2004].

For the CR1921, WSW-3DMHD did not predict the fast solar wind profile during April 10-17 which was caused by a MC crossing starting on April 11; neither did by Arge *et al.* [2004]. The $V_{GF}$ formula is modeled with quiet solar wind parameters and therefore it fails to predict solar wind disturbances caused by the crossing of the coronal mass ejection and its driven shock. To predict such a solar wind disturbance, a proper solar disturbance is required to add into the inner boundary of the simulation. In the following section, we will demonstrate the input requirement of solar disturbance for the solar wind condition.

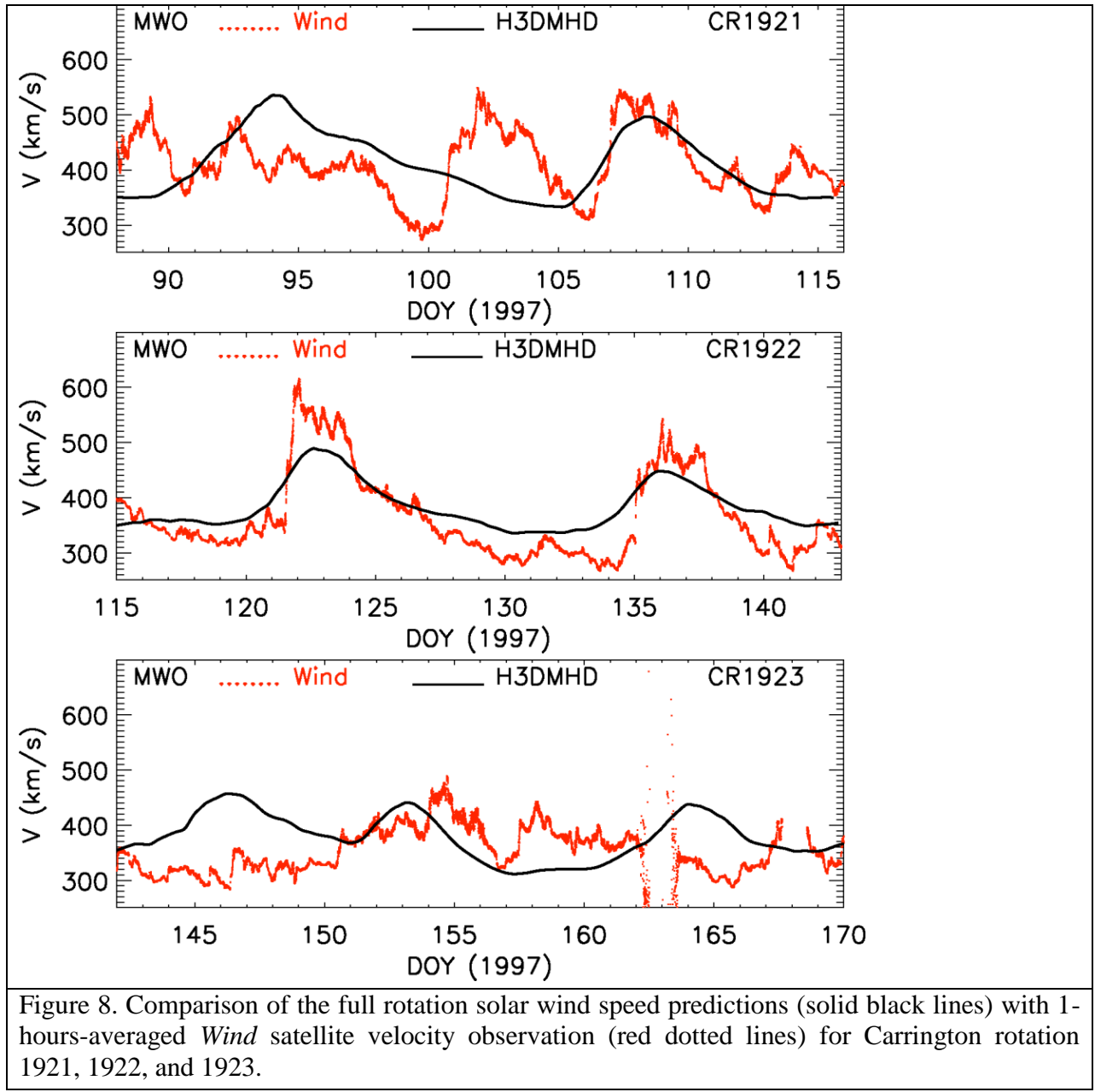

Figure 8. Comparison of the full rotation solar wind speed predictions (solid black lines) with 1-hours-averaged *Wind* satellite velocity observation (red dotted lines) for Carrington rotation 1921, 1922, and 1923.

**3.3 Validation of the good fit formula during non-quiet solar period**

In this Section we test the capability of $V_{GF}$ formula in solar active periods and the effect of solar disturbance (e.g., CME and its driven shock) on the solar wind profile. Two CMEs that occurred in September 2017 are simulated. Many solar activities (e.g., CMEs) were recorded in early September 2017. STEREO-A recorded two Sun-Earth directed CMEs, which occurred on the 2017-09-04 (referred as CME04) and 2017-09-06 (referred as CME06). The average CME propagating speed in the field of view (FOV) of STEREO-A for the CMEs occurred on the 4[th] and 6th was 866 km/s and 1308 km/s, respectively. A pressure pulse is inserted into the lower boundary of the simulation domain to simulate the CMEs.

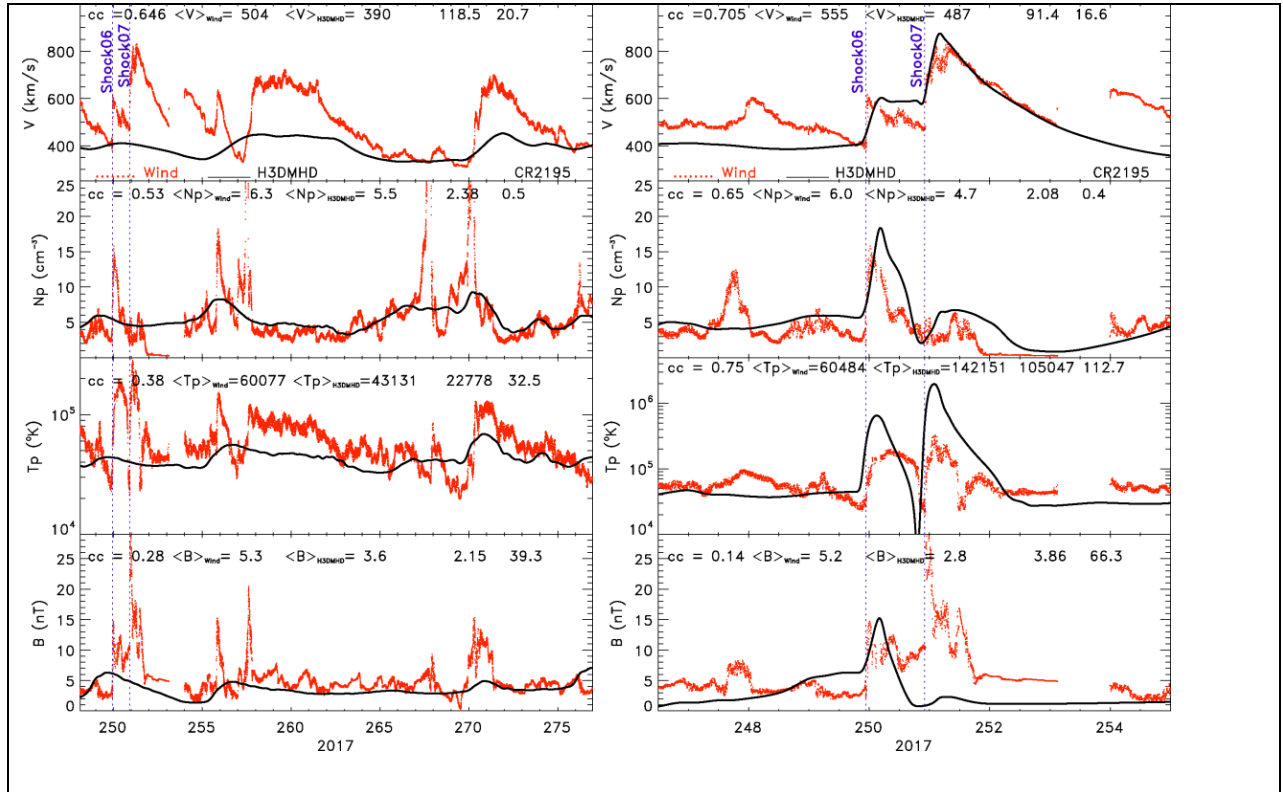

Figure 9. Comparison of solar wind speed, density, temperature, and temperature from the *WIND* spacecraft (red-dotted lines) with WS-H3DMHD prediction (black solid lines) during September-October 2017 without adding simulated CME perturbation (left panel), and during 04-11 September 2017 with two CMEs perturbation on 04-09-2017 and 06-09-2017 (right panel), respectively. Blue vertical dotted lines indicated the interplanetary (IP) shock arrival time at the *WIND* spacecraft. Shock06 and Shock07 represent the IP shock which arrived at the *WIND* on the 6[th] and 7[th] of September.

A comparison of the observed solar wind (speed, density, temperature, and magnetic field) with simulation without and with a CME perturbation input are showed in Figure 9A (left panel) and 9B (right panel) during 05/09/2017 – 03/10/2017, respectively For the case without a CME perturbation, values of the cc are 0.646, 0.53, 0.38, and 0.28 for *V, Np, Tp*, and *B*; and values of the MAPE are 20.7%, 0.5%, 32.5%, and 39.3% for *V, Np, Tp*, and *B*, respectively.

The simulated *Np, Tp*, and *B* match well with the basic trends of observation (see 2$^{nd}$, 3$^{rd}$, and 4$^{th}$ panels of Fig. 9A). However, the simulated velocity is far off of the observation (see top panel of Fig. 9A). Therefore, we conclude that G3DMHD is not able to predict the fast streams in September 2017. Figure 9A shows that the simulated undisturbed solar wind speed was 500 km/s slower than the observation between 05-09-2017 and 03-10-2017. All the high-speed solar wind streams are not predicted by the G3DMHD. One may suspect the prediction capability of WSW-3DMHD during the non-quiet solar period. Note that the V$_{GF}$ was introduced to reproduce background solar wind condition in the quiet period. STEREO-A had recorded two Sun-Earth-directed CMEs on the 4$^{th}$ and 6$^{th}$ of September, 2017. Perturbations of these two CMEs were inserted into the lower boundary of the WSW-3DMHD and the results were presented in Figure 9B.

Figure 9B shows a similar comparison as Figure 9A but with pressure pulse perturbations in the simulation. The values of cc are 0.705, 0.65, 0.75, and 0.14 for *V, Np, Tp*, and *B*, respectively. The values of MAPE are 16.6%, 0.4%, 112.7%, and 66.3% for V, Np, Tp, and B, respectively. The values of cc and MAPE are 0.705, 0.65, 0.75, 0.14; and 16.6%, 0.4%, 112.7%, 66.3% for V, Np, Tp, and B, respectively. The two vertical blue dotted lines in Figure 9B indicate the arrival time of interplanetary shocks at the *Wind* spacecraft on 06-09-2017 (referred to Shock06) and 07-09-2017 (referred to Shock07). The simulated solar wind speed at both upstream and downstream of Shock06 matches very well with the observation (see top panel of Fig. 9B). The simulated upstream speed of Shock07 is slightly higher than the observation, but the simulated downstream speed of Shock07 matches very well with the observation for about two days. The value of B at the downstream of Shock06 matches very well with the observation, but is poor for Shock07. A poor simulation result of B both upstream and downstream of Shock07 may be due to the fact that our simulation does not contain flux-rope structure, a very common problem in most data-driven global MHD models. Dynamic pressure pulses are often used to simulate the perturbation of CMEs [*e.g.,* Odstricil *et al.* 2005; Wu *et al.* 2007].

The about simulation result shows clearly that $V_{GF}$ is capable of reproducing the background solar wind in quiet solar periods. When there are solar events, such as CMEs, additional plasma perturbations are required at the inner boundary. Further investigation is needed to confirm the capability of the $V_{GF}$ formula for the long-term studies and CME events

## 4. Conclusions and Remarks

In the present study, we have demonstrated a computational scheme that derives the background solar wind speed at 18 solar radii. This scheme employs the conservation of mass, conservation of magnetic flux tube, and Bernoulli's principle in conjunction with the expansion factor derived from the Wang and Sheeley [1990] algorithm to model the solar wind speed with the formula $V_{18Rs} = V_1 + V_2 f_s^{\alpha}$. A set of the three parameters were tested to provide the inner boundary values for the MHD simulation we performed 216 simulations for CR2082 and found the best choice for the three parameters were $V_1 = 200\pm50$, $V_2 = 500\pm100$, and $a = -0.4$. Based on the results of single Carrington rotation, capabilities of the good fit formula, $V_{GF} = 150 + 500 f_s^{-0.4}$ was also validated in different solar cycles/activities, *i.e.*, in the years of 1995, 1997, 2004, 2009, and 2017. The $V_{GF}$ was found to be capable of being used in different solar activity, or/and solar cycles. To improve the accuracy of the prediction for the solar wind condition at 1 AU, a CME perturbation has to be added into the simulation if there is any.

In this study, we also compared our results with previous studies [Arge *et al.* 2000; 2004]. Comparisons between two models (WSA and WSW-3DMHD) are listed as following: a) Results of using $V_{GF}$ as input to drive G3DMHD model is better than the results of WSA using the full rotation (FR), or daily updated (DU). b) WSA using the modified daily updated (MDU) 4-day-advanced solar wind speed predictions is slightly better than that for WSW-3DMHD. c) Results of using $V_{GF}$ as input to drive 3DMHD model is better than the WSA formula. The present study put in doubt the use of an extra parameter (*i.e.,* the angular width from the nearest coronal hole).

While the empirical formula is derived using our G3DMHD model (used briefly as mentioned earlier for WSW+3DMHD), the result could be used for other similar MHD models with little to no change. This could be an interesting topic for future study. Combed with the empirical formula, some conservation laws, and the G3DMHD model, it can provide a powerful tool for space weather forecasting. In this study, several Carrington rotations were investigated and a couple of CME events were studied. A long-term study and/or a study with one and/or more CME events can definitely improve the validation work and will be addressed in the future.

## 6. Acknowledgment

All data used in this study are obtained from the public domain. We thank the OMNI PI teams and the National Space Science Data Center at Goddard Space Flight Center, National Aeronautics and Space Administration for management and providing solar wind plasma and magnetic field data. This study is supported partially by the Chief of Naval Research (CCW, YMW). The work of K.L. was supported by NASA grant NNX14A83G to the Johns Hopkins University Applied Physics Laboratory. The authors thank Dr. Christopher Kung from Engility/DoD High Performance Computing Modernization Office PETTT program for his technical assistance in parallelizing the G3DMHD code.

Improvements to the HAF solar wind model for space weather predictions, *J. Geophys. Res. 106(A10),* 20985-21002, 2001.

Gopalswamy, N/. P. Makela, H. Xie, S. Akiyama, and S. Yashiro, CME interactions with coronal holes and their interplanetary consequences, J. Geophys. Res. 114, A00A22, doi: 10.1029/2008JA013686, 2009.

Gosling, J.T., E. Hilder, R.M. MacQueen, R.H. Munro, A.I. Poland, C.L. Ross, Direct observations of a flare related coronal and solar wind disturbance, *Solar Physics* **40**, 439-448, 1975.

Gosling, J.T., 1990, in C.T. Russell, E.R. Priest, and L.C. Lee (eds.), Physics of Magnetic Flux Ropes, *Geophys. Monogr.,* **58**, Washington, DC, p.343., 1990.

Groth, Clinton P. T.; De Zeeuw, Darren L.; Gombosi, Tamas I.; Powell, Kenneth G., Global three-dimensional MHD simulation of a space weather event: CME formation, interplanetary propagation, and interaction with the magnetosphere, *Journal of Geophysical Res.* **105**, 25053-25078, 2000.

Han, S. M., A numerical study of two dimensional time-dependent magnetohydrodynamic flows. In: *Ph.D. Thesis*, University of Alabama in Huntsville, 1977.

Han, S. M., S. T. Wu, and M. Dryer, A three-dimensional, time-dependent numerical modeling of super-sonic, super-Alfvénic MHD flow, *Computers and Fluids*, **16,** 81-103, 1988.

Hayashi, K., M. Tokumaru, K. Fujiki, and M. Kojima (2011), Three-dimensional solar wind structures obtained with MHD simulation model using observation-based time-varying inner boundary map paper presented at 10th ASP Conference, *Astron. Soc. of the Pac.*, Kaanapali, Hawaii, 13–18 March, 2011.

Hess, P., and R.C. Colaninno, Comparing Automatic CME Detections in Multiple LASCO and
29